\documentclass[12pt]{article}
\usepackage{graphicx,amssymb,amsfonts}
\newlength{\extraspace}
\newlength{\extraspaces}
\newcommand{\be}{\begin{equation}
\addtolength{\abovedisplayskip}{\extraspaces}
\addtolength{\belowdisplayskip}{\extraspaces}
\addtolength{\abovedisplayshortskip}{\extraspace}
\addtolength{\belowdisplayshortskip}{\extraspace}}
\newcommand{\ee}{\end{equation}}

\newcommand{\ba}{\begin{eqnarray}
\addtolength{\abovedisplayskip}{\extraspaces}
\addtolength{\belowdisplayskip}{\extraspaces}
\addtolength{\abovedisplayshortskip}{\extraspace}
\addtolength{\belowdisplayshortskip}{\extraspace}}
\newcommand{\ea}{\end{eqnarray}}

\begin{document}

\begin{center}
{\bf Teleparallel equivalent of general relativity and local Lorentz transformation: Revisited}\footnote{ PACES numbers: 04.50. Kd, 04.70.Bw, 04.20. Jb\\
\hspace*{.5cm}
 }
\end{center}
\begin{center}
{\bf Gamal G.L. Nashed$^{1,2,3}$ and B. Elkhatib$^2$}
\end{center}

\bigskip
$^{1}$Centre for theoretical physics, the British University in Egypt, 11837 - P.O. Box 43, Egypt.\vspace{0.2cm}\\
$^{2}$Mathematics Department, Faculty of Science, Ain Shams University, Cairo, Egypt.\vspace{0.2cm}\\
 $^{3}$Egyptian Relativity Group (ERG).

\bigskip
\centerline{ e-mail:nashed@bue.edu.eg}

\hspace{2cm} \hspace{2cm}
\\
\\
\\
\\

It is well known that the field equations of  teleparallel theory which is equivalent to general relativity (TEGR) completely agree  with the field equation of general relativity (GR). However, TEGR has six extra degrees of freedom which spoil the true physics. These extra degrees are related to the local Lorentz transformation. In this study, we give three different tetrads of flat horizon space-time that depend only on the radial coordinate. One of these tetrads contains an arbitrary function which comes from local Lorentz transformation. We show by explicate calculations that this arbitrary function spoils the calculations of the conserved charges. We formulate {\it a skew-symmetric tensor} whose vanishing value put a constraint  on the arbitrary function. This constraint  makes the conserved charges are free from the arbitrary function.
\section{\bf Introduction}

In the theory of GR, the  gravitational field is gained by the curvature of space-times. Particles are obliged by the curvature of the space-times to move on geodesics. Therefore, the theory of GR is indeed geometrized by the gravitational field. The theory of GR has some constraints on the classical level. The forecasts of GR are in agreement with the experimental data accessible till now. Unification of
the main four  forces has continued as a favorite subject in physics. Because of the unknown attitude of GR at the quantum level, the scientistic community  required the unified theory. For this purpose, Einstein did his  attempt  which was failed. In this research, we study his suggested theory, which is called TEGR  as an alternative theory of gravitational field in the absence of unifying it with quantum theory.

 The TEGR  is established on Weitzenb\"{o}ck geometry \cite{Wr}. In this theory torsion behaves as a force on objects. Therefore, in TEGR, there is no geodesics but only  force equations \cite{AP5}. The GR theory is characterized mathematically by Einstein's field equations, in which  the geometry of the space-time is described by one side and
 the physics of the space-time is described by the other side.  To find a solution in GR is not an easy task, therefore, we demanded certain symmetry constraints on the space-time metric. The significance of   symmetries is   understandable in GR owing to the laws of
conservation of   matter in the space-time which can be studied and realized with the aid of the  symmetry constraints \cite{Pa6}.

The TEGR theory is identical to GR. The GR theory is a geometric theory constructed on the Levi-Civita connection, that has curvature and zero torsion.  However, TEGR employs anther  connection. This connection is given by Weitzenb\"{o}ck which is  defined on a space so that is globally flat, i.e., has zero curvature.  This connection is known as a Weitzenb\"{o}ck connection and  has a vanishing Ricci tensor but non-trivial torsion.  This connection is employed to construct a Lagrangian  built on a gravitational scalar named as the torsion scalar, $T$. The dynamics of this Lagrangian are equal to GR and this follows from the result
\[R=-T-B,\] with $R$ being  Ricci scalar and $B$ is a boundary term linked to the divergence of the torsion. Due to the fact that $B$ is a total
derivative, it does not contribute to the equations of motion and therefore, the Lagrangian of TEGR is equal to that of GR.

The aim  of  this study is to discuss the effect of the extra degrees of TEGR on the true physics and how one can fix  these degrees so that they do not contribute to the true physics. In \S 2, an introduction to TEGR theory is presented. In \S 3, several tetrad fields having flat horizons are given and application  to the field equations of TEGR is explained. New analytic, solutions are derived in \S 3. In \S 4,  calculation of the conserved quantities of each solution have been carried out and we have shown how the unphysical extra degrees contribute  to the true physics.  In \S  5, we gave a skew-symmetric tensor which constrains the extra degree and shows the effect of this tensor on the acceleration components of an observer  at infinity.  Main results are discussed in final section.
\newpage
\section{Introduction to TEGR theory}
Teleparallel theory equivalent to general relativity considers as another construction of GR of Einstein. The main entity of TEGR theory is the tetrad fields\footnote{The Greek symbols indicate the elements of tangent
space  and Latin components
indicate the symbols of the spacetime.} $(e^{i}{_{\alpha}})$. In TEGR theory, the metric can be built using the tetrad: ${\rm g}_{\alpha \beta}=\lambda_{i j} {\rm e}^{i}{_\alpha} {\rm e}^{j}{_\beta}$, with $\lambda_{i j}=\textmd{diag}(1,-1,-1,-1)$ being the Minkowskian metric, thus the symmetric  connection , Levi-Civita , ${\mathring{\rm \Gamma}}{^{\alpha}}{_{\beta \gamma}}$ can be constructed. Nevertheless, it is likely  to build Weitzenb\"{o}ck non-symmetric connection ${\Gamma^\alpha}_{\beta \gamma}={e_i}^{\alpha} \partial_{\gamma} {e^i}_{\beta}$ \cite{Wr}. The Weitzeinb\"{o}ck  spacetime is labled as a pair $(M,e_{\mu})$, whereas $M$ is a $D$-dimensional  manifold and $e_{\beta}$ ($\beta=0,\cdots, 3$) are $D$-linear independent vector  defined globally on $M$. The covariant derivative of the tetrad using Weitzenb\"{o}ck connection  is  vanishing,  i.e. $\nabla_{_{\alpha}}{\rm e}^{i}{_{\beta}}\equiv 0$. Therefore, the vanishing of the covariant derivative of the tetrad  identifies the auto parallelism or absolute parallelism restriction. Actually, the operator $\nabla_{\alpha}$ has a big issue, that is not invariant under  local  Lorentz transformations (LLT). The issue   permits all LLT invariant quantities to rotate freely at  each point of the space \cite{M2013}. Thus, the symmetric metric  is not able to guess one form of tetrad field; therefore, the  additional degrees of freedom  have to be restricted such that  one physical frame can be identified.  The Weitzenb\"{o}ck connection is has a vanishing curvature however it has a  torsion defined by
\begin{equation}\label{torsion}
{\rm T}^{\alpha}{_{\beta \gamma}} :={\rm \Gamma}^{\alpha}{_{\beta \gamma}}-{\rm \Gamma}^{\alpha}{_{\gamma \beta}}.
\end{equation}
The contortion is defined as
\begin{equation}\label{contortion}
{{\rm K}^{\alpha \beta}}_\gamma :=-\frac{1}{2}\left({{\rm T}^{\alpha \beta}}_\gamma-{{\rm T}^{\beta \alpha }}_\gamma-{{\rm T}_\gamma}^{\alpha \beta}\right).
\end{equation}
In the TEGR  one can construct three Weitzenb\"{o}ck invariants: $I_{1}=T^{\alpha \beta \gamma}T_{\alpha \beta \gamma}$, $I_{2}=T^{\alpha \beta \gamma}T_{\beta \alpha \gamma}$ and ${\rm I}_{3}={\rm T}^{\alpha}{\rm T}_{\alpha}$, where ${\rm T}^{\alpha}={\rm T}_{\beta}{^{\alpha \beta}}$. We next define the invariant ${\rm T}={\rm A}_1{\rm I}_{1}+{\rm A}_2{\rm I}_{2}+{\rm A}_3{\rm I}_{3}$, where ${\rm A}_k$, $k=1,2,3$  are  constants \cite{M2013}. When ${\rm A}_1=1/4$, ${\rm A}_2=1/2$ and ${\rm A}_3=-1$  the invariant ${\rm T}$ will be identical with Ricci scalar ${\rm R}^{(\mathring{\Gamma})}$, up to divergence term.  In this sense, the  teleparallel  gravity will be identical with GR.   The torsion scalar of TEGR  is defined as
\begin{equation}\label{Tor_sc}
{\rm T} := {{\rm T}^\alpha}_{\beta \gamma}{{\rm S}_\alpha^{\beta \gamma}},
\end{equation}
with ${{\rm S}_\alpha}^{\beta \gamma}$ being the superpotential tensor defined as
\begin{equation}\label{Stensor}
{{\rm S}_\alpha}^{\beta \gamma}:= \frac{1}{2}\left({{\rm K}^{\beta \gamma}}_\alpha+\delta^\beta_\alpha {{\rm T}^{\rho \gamma}}_\rho-\delta^\gamma_\alpha {{\rm T}^{\rho \beta}}_\rho \right).
\end{equation}
The tensor ${{\rm S}_\alpha}^{\beta \gamma}$ is skew symmetric tensor in the last two indices.

The identity between torsion scalar and Ricci one is given by
\begin{equation}\label{Bianchi1}
{\rm R}^{(\mathring {\Gamma})}=-{\rm T}^{(\Gamma)}-2\ {\nabla}_{\alpha}{\rm T}^{\beta \alpha}{_{\beta}},
\end{equation}
 such that the covariant derivative is with regard to the symmetric  connection, i.e. Levi-Civita. The  total derivative term is the one on the second term on the right-hand side of the above equation. Therefore, there will be no contribution of the divergence term to the variation of the scalar torsion $T$ instead of the Ricci scalar. In this sense, the  torsion and Ricci  scalars are identical. Despite the quantitative equivalence, they, $T$ an $R$,  are qualitatively not equivalent. For instant,  Ricci scalar  tensor is invariant under LLT whilst the divergence term ${\nabla}_{\alpha}{\rm T}^{\beta \alpha}{_{\beta}}$ is not invariant and therefore  the torsion scalar. Therefore, the theory of TEGR \textit{action} is not form invariant with respect to LLT \cite{1010.1041,1012.4039,KS2015}.

The  action  of the gauge gravitational field Lagrangian is given by \cite{M94,M2002}
\begin{equation}\label{action}
\mathcal{{\rm S}}=\frac{{\rm M}_{\textmd{\tiny Pl}}^2}{2}\int |e| \left[({\rm T}-2\Lambda)+ \mathcal{{\rm L}}_{M}(\Phi_{A})\right]~d^{4}x,
\end{equation}
with $\mathcal{{\rm L}}_{M}$ being the Lagrangian of matter fields $\Phi_{A}$ and $M_{\textmd{\tiny Pl}}$ being  the   mass of Planck, that is connected to gravitational constant $G$ through  ${\rm M}_{\textmd{\tiny Pl}}=\sqrt{\hbar c/8\pi G}$. In this study we use the units  $G = c = \hbar = 1$ and  $|e|=\sqrt{-g}=\det\left({e^i}_\alpha\right)$. Making variation of Eq. (\ref{action}) regard to the tetrad fields ${{\rm e}^i}_\alpha$ give the following field equations \cite{M94}
\begin{equation}\label{TEGR}
\partial_{\beta}(e {\rm S}_{i }{^{\alpha \beta}})=4\pi {\rm e} {\rm e}_{i}{^\beta}({\rm t}_{\beta}{^{\alpha}}+{\rm \Theta}_{\beta}{^{\alpha}}),
\end{equation}
with ${\rm S}_{i}{^{\alpha \beta}}={\rm e}_{i}{^\rho}{\rm S}_{\rho}{^{\alpha \beta}}$, being the (pseudo) tensor ${\rm t}_{\alpha}{^\beta}$ and
\begin{equation}\label{grav_EM_tensor}
{\rm t}_{\alpha}{^{\beta}}=\frac{1}{16\pi}[4{|rm T}^{\rho}{_{\gamma \alpha}}{\rm S}_{\rho}{^{\beta \gamma }}-\delta^{\beta}_{\alpha}({\rm T}-2{\rm \Lambda})],
\end{equation}
is the  energy-momentum tensor
\begin{equation}\label{EM_tensor}
{\rm \Theta}_{\alpha}{^\beta}={\rm e}^{i}{_{\alpha}}\left(-\frac{1}{e}\frac{\delta {\rm L}_{M}}{\delta {\rm e}^{i}{_{\beta}}}\right).
\end{equation}
As the tensor ${\rm S}_{i}{^{\alpha \beta}}$ is anti-symmetric, i.e ${\rm S}_{i}{^{\alpha \beta}}=-{\rm S}_{i}{^{\beta \alpha}}$, this leads to
$\partial_{\alpha}\partial_{\beta}({\rm e} {\rm S}_{i}{^{\alpha \beta}})\equiv 0$ \cite{M2013}. Therefore,
$$\partial_{\beta}\left[{\rm e} {\rm e}_{i}{^\rho}({\rm t}_{\rho}{^{\beta}}+{\rm \Theta}_{\alpha}{^{\beta}})\right]=0.$$
The pseudo-tensor ${\rm t}_{\alpha}^{\beta}$ is disappeared in the theory of GR. To prob its behavior  we see that the previous equation leads us to the conservation 
$$\frac{\rm d}{\rm dt}\int_{V}{\rm e} {\rm e}_{i}{^\alpha}({\rm t}_{0\, \alpha}+{\rm \Theta}_{0\, \alpha})d^3 x=-\oint_{\Sigma}\left[{\rm e} {\rm e}_{i}{^\alpha}({\rm t}_{j  \alpha}+{\rm \Theta}_{j \alpha})\right]d\Sigma^{j}.$$
 The integration of the previous equation is  carried out on 3-dimensional volume limited by the surface. Therefore,  ${\rm t}_{\alpha}{^\beta}$ represent the energy-momentum tensor of the gravitational field \cite{M94}.
\section{Flat transverse solutions}
We apply the TEGR field equations (7) to the {\it first flat transverse section}   which gives  the following tetrad  written in  cylindrical coordinate ($t$, $r$, $\phi$, $z$) as:
\begin{eqnarray}\label{tetradf}
\nonumber \left({e^{i}}_{\mu}\right)=\left(
 \begin{array}{cccc}
    {\rm a(r)} & 0 & 0 & 0 \\
    0&{\rm b(r)} &0 & 0 \\[5pt]
    0&0&r&0\\[5pt]
    0&0&0&r \\[5pt]
  \end{array}
\right),&\\
\end{eqnarray}
Substituting from (\ref{tetradf}) into (\ref{Tor_sc}) we calculate  torsion scalar as
\begin{equation}\label{Tsc}
    T=-\frac{2(2a' r+a)}{r^2ae^2},
\end{equation}
where $a':=\frac{da(r)}{dr}$ .
Using  (\ref{tetradf})  in (7) we get
\begin{eqnarray}
& &  \Theta_0{}^0=\frac{\Lambda r^2e^3+2rb'-b}{2r^2e^3}, \quad  \Theta_1{}^1=\frac{\Lambda r^2e^2a-2ra'-a}{2r^2e^2a},\nonumber\\
& & Theta_2{}^2=\Theta_3{}^3=\frac{\Lambda re^3a-rba''+ra'b'+ab'-ba'}{2re^3a}. \nonumber\\
\end{eqnarray}
The solution of the above differential equation has the form
\begin{equation} a(r)=\frac{c_1\sqrt{\Lambda r^3+3c_2}}{\sqrt{r}}, \qquad \qquad b(r)=\mp\frac{\sqrt{3r}}{\sqrt{\Lambda r^3+3c_2}},\end{equation}

The {\it second  flat transverse section}  tetrad is given by
\begin{eqnarray}\label{tetrad1}
\nonumber \left({{e_1}^{i}}_{\mu}\right)=\left(
 \begin{array}{cccc}
    a(r) & 0 & 0 & 0 \\
    0&b(r)\cos\phi &-r\sin\phi & 0 \\[5pt]
    0&b(r)\sin\phi&r\cos\phi&0\\[5pt]
    0&0&0&r \\[5pt]
  \end{array}
\right).&\\
\end{eqnarray}
Tetrad (\ref{tetrad1}) is related to (\ref{tetradf}) by a rotation matrix given by
\begin{eqnarray}\label{tetrad}
\nonumber \left({\Lambda^{i}}_{j}\right)=\left(
 \begin{array}{cccc}
    1 & 0 & 0 & 0 \\
    0&\cos\phi &-\sin\phi & 0 \\[5pt]
    0&\sin\phi&\cos\phi&0\\[5pt]
    0&0&0&r \\[5pt]
  \end{array}
\right).&\\
\end{eqnarray}
Substituting from (\ref{tetrad1}) into (\ref{Tor_sc}) we evaluate the torsion scalar as
\begin{equation}\label{Tsc}
    T=-\frac{2(rba' +ab-2ra'-a)}{r^2ae^2}.
\end{equation}
Using  (\ref{tetrad1})  in (7) we get
 get the same  differential equation given by Eq. (12) which have the same solution given by Eq. (13).

The {\it third  flat transverse section}  tetrad is given by
\begin{eqnarray}\label{tetrad3}
\nonumber \left({e_3{}^{i}}_{\mu}\right)=\left(
 \begin{array}{cccc}
    -a(r)\mathfrak{L} & -\mathfrak{H}b(r)\cos\phi &  -r\mathfrak{H}\sin\phi  & 0 \\
    a(r) \mathfrak{H}\cos\phi &b(r)
    (\sin^2\phi+\mathfrak{L} \cos^2\phi )&r\sin\phi\cos\phi(\mathfrak{L}-1)  & 0 \\[5pt]
    a(r) \mathfrak{H}\sin\phi  &b(r)\sin\phi\cos\phi(\mathfrak{L}-1)&r
    (\cos^2\phi+\mathfrak{L} \sin^2\phi )&0\\[5pt]
    0&0&0&r \\[5pt]
  \end{array}
\right),&\\
\end{eqnarray}
where $\mathfrak{L}=\sqrt{1+\mathfrak{H^2}}$ and $\mathfrak{H}$ is an arbitrary function of $r$. This arbitrary function perseveres the flat horizon of tetrad (\ref{tetrad3}).
Tetrad (\ref{tetrad3}) is related to (\ref{tetradf}) by a LLT matrix given by
\begin{eqnarray}\label{tetrad5}
\nonumber \left({\Lambda_1{}^{i}}_{\mu}\right)=\left(
 \begin{array}{cccc}
    -\mathfrak{L} & -\mathfrak{H}\cos\phi & -\mathfrak{H}\sin\phi & 0 \\
    \mathfrak{H}\cos\phi &\sin^2\phi+\mathfrak{L}\cos^2\phi &\sin\phi \cos\phi(\mathfrak{L}-1)& 0 \\[5pt]
    \mathfrak{H}\sin\phi &\sin\phi \cos\phi(\mathfrak{L}-1) &\cos^2\phi+\mathfrak{L}\sin^2\phi & 0\\[5pt]
    0&0&0&1 \\[5pt]
  \end{array}
\right).&\\
\end{eqnarray}
Substituting from (\ref{tetrad3}) into (\ref{Tor_sc}) we evaluate the torsion scalar as
\begin{equation}\label{Tsc}
    T=\frac{2(ba\mathfrak{H}^2+ba+rb\mathfrak{H}^2a'+rba'-ab\mathfrak{L}-b\mathfrak{L}ra'+rab\mathfrak{H}\mathfrak{H}'-a\mathfrak{L}-2ra'\mathfrak{L}))}{r^2ae^2\mathfrak{L}}.
\end{equation}
Using  (\ref{tetrad3})  in (7) we get
the same  differential equation given by Eq. (12) which have the same solution given by Eq. (13).   Therefore, the field equations (7) are not able to give a specific form of the arbitrary function $\mathfrak{H}$ as expected due to the fact that the field equations (7) are equivalent to GR.  In the next section we are going to discuss the physical relevance of each tetrad.
\section{Conserved quantities and intuitive  of Einstein-Cartan theory }
There are many modifications of GR.  In the frame of  theoretical physics, the theory of Einstein-Cartan (EC), which is identified  as  ``Einstein-Cartan-Sciama-Kibble theory", is  classic  gravitation theory like GR  in which its   connection has no  skew symmetric part. Thus,  in EC, the torsion tensor can be accompanied  to the spin of  the matter, by the  same pattern  the curvature is joined to the   momentum and energy of the matter.  Actually,  spin of   matter using non-flat spacetime demands that  the  torsion  tensor does not vanishing  but  be  a variable, i.e., stationary action. The theory of  EC deals with   the  torsion tensor and metric as independent which  give  the right extension of  conservation law   in the existence  of the gravitational field. The EC   theory   constructed  by \'{E}lie Cartan in \cite{Ce} and recently it has many application \cite{Ce1}.
  The Lagrangian of EC has the form \cite{OR6}:\begin{equation}
{\rm {\cal L}}({\rm \vartheta}^i, \
{{\rm  \Gamma}^j}_{k})=-\frac{1}{2\kappa}\left({\rm R}^{i j}\wedge
\eta_{i j}-2{\rm \Lambda} \eta\right),\end{equation} with ${\rm \vartheta}^i$  being the co-frame one form, ${{\it \Gamma}^j}_{k}$ being the connection, $\kappa$  and ${\rm \Lambda}$  are the gravitational  and the cosmological constants. The Lagrangian  given by Eq. (20) is a form invariant under  diffeomorphism  and  Lorentz local transformation \cite{OR6}.  Carrying out the  principle of least  action to equation (20)  we get  \cite{OR6,Kw1} \begin{eqnarray} &  & {\rm   E}_{i}:= -\frac{1}{2\kappa}\left({\rm R}^{ j k}\wedge
\eta_{i j k}-2{\rm \Lambda} \eta_i\right) ,   \qquad \qquad {\rm H}_{i j}:=\frac{1}{2\kappa}\eta_{i j},\end{eqnarray}
where $\eta_{i j}$ is a two form given in the Appendix A, ${\rm   E}_{i}$ is the energy-momentum and ${\rm H}_{i j}$ is the rotational  gauge field momentum.
The  momentum of translation and the spin take the following form \begin{eqnarray} & &  {\rm H}_{i}:=-\frac{\partial
{\rm {\cal L}}}{\partial {\rm T}^i}=0, \qquad  \qquad {\rm E}_{i j}:= -{\rm \vartheta}_{[i}\wedge {\rm H}_{j]}=0.\end{eqnarray} The
minimally coupling of matter is supposed such that $\frac{\partial
{\rm L}_{Matter}}{\partial {\rm T}^{i}}=0$ and $\frac{\partial {\rm L}_{matter}}{
\partial {\rm R}^{i}_{j}}=0$.\footnote{ The  derivative of the coframe vanishes
if $\xi$ is a Killing vector field, i.e.  ${\cal L}_{\rm \xi} {\rm \vartheta}^i$= 0 \cite{OR6}.}  The conserved current is given by  \cite{OR6}
\begin{eqnarray} & & \jmath[\xi]=\frac{1}{2\kappa}d\left\{^{^{*}}\left[d {\it k}+\xi\rfloor\left({\it \vartheta}^i\wedge
{\rm T}_j\right)\right]\right\}, \qquad \textrm{where}\nonumber\\
& & \nonumber\\
& & {\rm k}={\rm \xi}_i
{\rm \vartheta}^i, \qquad \textrm{and} \qquad {\rm \xi}^i={\rm \xi}\rfloor{\rm \vartheta}^i, \end{eqnarray}
where $*$ is defined as the Hodge duality and ${\rm \xi}$ is an arbitrary vector
field ${\it \xi}={\rm \xi}^i\partial_i$.  ${\rm \xi}^i$ in this study  are four  parameters ${\it \xi}^0$, ${\it \xi}^1$, ${\it \xi}^2$
and ${\it \xi}^3$.
Because this study is in the frame of  theory of TEGR which is  equivalent to GR, thus, torsion is nil
  and   conserved charge,  Eq. (23), is given by
\begin{equation} {\rm {\cal {\rm Q}}}[{\rm \xi}]={\rm \frac{1}{2\kappa}\int_{\partial {\rm S}}{^*}d{\rm k}}. \end{equation}
Expression (24)   was given  by   Komar  \cite{Ka2}--\cite{Ka3} and is invariant
under  diffeomorphism.

  The coframe ${\it \vartheta_1}^i$ of
solution (13) using tetrad (10)  has the form: \begin{equation} {{\it \vartheta}}^{\hat{0}} ={a} dt,\qquad \quad
{{\it \vartheta_1}}^{\hat{1}}=b dr, \qquad { {\it \vartheta}_1}^{\hat{2}}= rd\theta, \qquad
 {\it \vartheta_1}^{3}=rd\phi.  \end{equation}

By using equation
(25) in equation  (24) we obtain \begin{eqnarray} & &   {\it  k}=a^2{\it \xi}_0dt - {b}^2{\it \xi}_1dr-r^2[{\it \xi}_2d\theta  +{\it \xi}_3d\phi].\end{eqnarray}
Total derivative of Eq. (26) gives
 \begin{equation} d{\rm  k}=2 [a a'  {\it \xi}_0 (dr \wedge dt)+r{\it \xi}_2 (d\theta \wedge dr)-r{\it \xi}_3 (dr \wedge d\phi)].\end{equation}
 From the inverse of (25)
   using (27) in (24) and taking
the Hodge-duality to $d {\rm k}$,  we get
\begin{equation} {{\cal {\rm Q}}}[{\rm \xi}_t]= {{\cal {\rm Q}}}[{\rm \xi}_r]={{\cal {\rm Q}}}[{\rm  \xi}_\theta]=
{{\cal {\rm  Q}}}[{\rm \xi}_{\phi}] =0.\end{equation}
  Equation (28) indicates in clear way that the total conserved charge  of
(13)  are nil. Carrying out the same procedure to tetrads (\ref{tetrad1}) and  (\ref{tetrad3})  we get the same result of Eq. (28).
Thus, equation (24) must redefined to get a well defined value, i.e.  Eq. (24) needs a  regularization.
\section{Regularization using relocalization}
Expression (24) is form invariant under  diffeomorphism and Lorentz   local  transformation. However, it is demonstrated   that plus to diffeomorphism and Lorentz   local  transformation there exists other  defect in the  form of conserved quantities. This defect lies in    equations of motion which  permit  a relocalization  \cite{OR6}. Therefore,  conserved quantities  can be altered using  relocalization. Relocalization  appeared from  amercement of  the gravitational Lagrangian through a total
derivative term has the form \begin{equation} {\rm S}'={\rm S}+d\Phi, \qquad \textrm{where} \qquad
{\it \Phi}={\rm \Phi}({{{{\it \vartheta}^{{}^{{}^{\!\!\!\!}}}}{_{}{_{}{_{}}}}}^{i}},
{{\it \Gamma}_i}^j, {\rm T}^i, {{\it R}_i}^j),\end{equation} with  ${{\it \Gamma}_i}^j$ is a  1-form  connection. The second term  in Eq. (29), i.e.
$d{\rm \Phi}$, amendments  the boundary part of the Lagrangian, premitting the equations of motion to be in a covariant form
( \cite{OR6} and references therein). It has been  explained that  the total conserved
charges could be regularized by applying a relocalization procedure.
It has been explained that to solve the odd result
derived in Eq. (28), one has to employ  relocalization given by  the boundary expression that appears in the Lagrangian. We use the relocalization in the form 
\[ {\it H}_{i j}\rightarrow {\it H}'_{i j}={\it  H}_{i j}-2\beta{\it  \eta}_{i j k l}{\it  R}^{k l},\] that  is originated from  the modification of the Lagrangian  \cite{OR6}
\begin{equation} {\it L}\rightarrow {\it S}'={\it S}+{\it \beta} d{\it \Phi},\] where \[ {\it H}'_{i j}=\left(\frac{1}{2\kappa}-\frac{4{\it \beta} { \Lambda}}{3}\right){\it \eta}_{i j}-2{\it \beta} {\it \eta}_{i j k l}\left({\rm R}^{k l}-\frac{\Lambda}{3}{{\it \vartheta}}^{k}{{\it \vartheta}}^{l}\right).\end{equation} Assuming ${\it \beta}$, which exists in  Eq. (30) has the value $\frac{3}{8\Lambda \kappa}$ in 4-dimension to confirm the cancelation of the vanishing value  (that comes from inertia) which exists in Eq. (28). Thus, the conserved charges after regularization has the form  \cite{OR6,OR7,OR8}
\begin{equation} {{\cal {\it J}}}[\xi]=-\frac{3}{4\kappa \Lambda }\int_{\partial S}
\eta_{i j k l}\Xi^{i j} {\it W}^{k l}, \end{equation} with
${\rm W}^{i j}$ is the Weyl 2-form described by \begin{equation} {\rm W}^{i j}=\frac{1}{2}{{\it C}_{k l}}^{i j}{{\it \vartheta}}^{k}\wedge {{\it \vartheta}}^{l},\end{equation} where ${C_{i j}}^{k l}={e_i}^\mu {e_j}^\nu {e^k}_\alpha {e^l}_\beta {C_{\mu \nu}}^{\alpha \beta}$ is the Weyl tensor and  $\Xi^{i j}$
is denoted by \begin{equation} \Xi_{i j}=\frac{1}{2}e_j\rfloor
e_i \rfloor dk.\end{equation}  The conserved
currents ${{\cal {\it J}}}[\xi]$  given by Eq (31) are form invariant under  diffeomorphism and Lorentz  local transformation. These
currents ${{\cal {\it J}}}[\xi]$ are linked to the vector field ${\rm \xi}$ on the
spacetime manifold.

 calculating the necessary components of Eq. (31) we get  \begin{eqnarray} {\it \Xi}_{01} &= -& \frac{a' \xi_0}{b}, \qquad \qquad {\it \Xi}_{13} =-\frac{\xi_3}{b}. \end{eqnarray}

  Using Eq. (34), we get the value of $\eta_{i j k l}{\rm \Xi}^{i j}
 {\it W}^{k l}$ in 4-dimension in the form \begin{eqnarray}  \eta_{i j k l}{\rm \Xi}^{i j}
 {\it W}^{k l}=\frac{2\sqrt{3} c_1c_2\xi_0(2\Lambda r^3-3c_2)}{3r^3}.\end{eqnarray} Substituting Eq. (35) in
(31) we get  \begin{equation} {{\cal
{\it J}}}[\xi_t]=\frac{\sqrt{3}}{2}\xi_0 \pi c_1c_2+\left(\frac{1}{r^3}\right),\qquad {{\cal
{\it J}}}[\xi_r]={{\cal
{\it J}}}[\xi_z]={{\cal
{\it J}}}[\xi_{\phi}]=0.\end{equation} Equation (36) shows that the two constants $c_1$ and $c_2$   may take the values $c_2=\frac{2}{\sqrt{3}\pi}$ and $c_2= M$ in which  total mass and angular momentum takes the form \cite{LZ, Dm}
\begin{equation} E=M+\left(\frac{1}{r}\right), \qquad \qquad {{\cal
{\it J}}}[\xi_r]={{\cal
{\it J}}}[\xi_z]={{\cal
{\it J}}}[\xi_{\phi}]=0.\end{equation}

Repeat the same calculation we get the non-vanishing components   $\Xi^{i j}$  of tetrad (14) to have the form \begin{eqnarray} {\it \Xi}_{01} &=& \frac{a' \cos\phi \xi_0}{b}, \qquad \qquad  {\it \Xi}_{02}=-\frac{a' \sin\phi \xi_0}{b}, \qquad      {\it \Xi}_{13}=-\frac{ \cos\phi \xi_3}{b}, \qquad      {\it \Xi}_{23}=-\frac{ \sin\phi \xi_3}{b}.\nonumber\\
& & \end{eqnarray}

  Using Eq. (31), we get the value of $\eta_{i j k l}{\rm \Xi}^{i j}
 {\rm W}^{k l}$ in 4-dimension in the  same form of Eq. (35) which gives the same conserved quantities as given by Eq. (37).

 %
   %

The survive  components   $\Xi^{i j}$  of tetrad (17) are
 \begin{eqnarray} {\it \Xi}_{01}&=& -\frac{  \xi_0 a'  [ (2\mathfrak{H}^2 +1)\cos^2\phi+\mathfrak{L}\sin^2\phi] }{b}, \quad   {\it \Xi}_{02}=-\frac{  \xi_0 a' \sin\phi \cos\phi(2\mathfrak{H}^2 +1-\mathfrak{L})}{b}, \qquad   {\it \Xi}_{03}=-\frac{  \xi_3 \cos\phi \mathfrak{H}}{b},\nonumber\\
 {\it \Xi}_{12}&=&\frac{  \xi_0 a'   \mathfrak{H}  \sin\phi (2[\mathfrak{L}-1]\cos^2\phi +1}{b}, \quad  {\it \Xi}_{13}=\frac{  \xi_3(\sin^2\phi+\cos^2\phi \mathfrak{L})}{b}, \qquad {\it \Xi}_{23}=\frac{  \xi_3\sin\phi\cos\phi(\mathfrak{L}-1)}{b}.\end{eqnarray}
 Using Eq. (31), we get the value of $\eta_{i j k l}{\rm \Xi}^{i j}
 {\it W}^{k l}$ in 4-dimension in the form \begin{eqnarray} \eta_{i j k l}{\rm \Xi}^{i j}
 {\it W}^{k l}=\frac{2\sqrt{3} c_1c_2\xi_0(2\mathfrak{H}^2[\mathfrak{L}-1]\sin^2\phi \cos^2\phi-2\mathfrak{H}^2-1)(2\Lambda r^3-3c_2)}{3r^3}.\end{eqnarray} Substituting Eq. (40) in (31) we  get \begin{equation} {{\cal
{\it J}}}[\xi_t]=\frac{\sqrt{3} c_1c_2\xi_0(2\mathfrak{H}^2[\mathfrak{L}-1]\sin^2\phi \cos^2\phi-2\mathfrak{H}^2-1)(2\Lambda r^3-3c_2)}{4r^3},\qquad {{\cal
{\it J}}}[\xi_r]={{\cal
{\it J}}}[\xi_\theta]={{\cal
{\it J}}}[\xi_{\phi}]=0.\end{equation} Equation   (41) shows that the arbitrary function  $\mathfrak{H}$ which describes inertia contributes to the true physics.

\section{Physical constrains on the inertia}
As we discussed in the previous sections that we have three tetrad fields reproduce the same metric.
The TEGR field equations  give the same solution of the two unknown  functions however, they can not able to give any specific form of the arbitrary function. Also we have shown that the scalar torsion depend on the tetrad as is known in the literature which means that it is not local Lorentz transformation. In the previous section we try to calculate the conserved quantities and show that they depend on the inertia. In this section, we are going to assume  specific form of skew-tensor and see if this tensor will help in solving the above problem or not?
It is well known that the field equations of $f(T)$ is non-symmetric \cite{Nprd}--\cite{WY13}\footnote{We will denote the symmetric part by ( ), for example, $A_{(\mu \nu)} = (1/2)(A_{\mu \nu} +A_{\nu \mu})$ and
the antisymmetric part by the square bracket [ ], $A_{[\mu \nu]} = (1/2)(A_{\mu \nu} -A_{\nu \mu})$.}
 \begin{eqnarray} & & S_{(\mu  \nu)}{}^\rho T_{,\rho} \
f(T)_{TT}-\left[e^{-1}{e^a}_\mu\partial_\rho\left(b{e_a}^\alpha
{S_\alpha}^{\rho \nu}\right)-{T^\alpha}_{\lambda \mu}{S_\alpha}^{\nu
\lambda}\right]f(T)_T-\frac{1}{4}g_{\nu \mu} (f(T)-2\Lambda)=4\pi
{\cal T}_{\nu \mu},\nonumber\\
& &  S_{[\mu  \nu]}{}^\rho T_{,\rho}  \  f(T)_{TT}=0.
\end{eqnarray}
Therefor, in TEGR the skew-symmetric is satisfied automatic due to the fact that $f_{TT}=0$.  Now let us check if the skew symmetric tensor
\begin{equation}
S_{[\mu  \nu]}{}^\rho T_{,\rho}, \end{equation} is vanishing for the above tetrad fields or not?
For the first tetrad given by Eq. (10) Eq. (43) is satisfied automatic. Therefore, tetrad (10) we call it a physical tetrad. For the second tetrad field, given by (14), Eq. (43) is satisfied identically.
%
%

 Using Eq. (43), we get for the third tetrad field,  Eq. (17),  the following non-vanishing components:
\begin{eqnarray} & &  S_{[1 0] 1} T^{,1}=\frac{\mathfrak{H}\cos\phi}{2\mathfrak{L}^{3/2}a b r^4}\Biggl(b\Biggl[r^2ab \mathfrak{L}^4a''+r^2\mathfrak{H}a^2b \mathfrak{L}^4 \mathfrak{H}''-r^2b \mathfrak{L}^4a'^2-raa'\mathfrak{L}^2(r\mathfrak{L}^2b'-b [ r\mathfrak{H}\mathfrak{H}'-\mathfrak{L}^2] )\nonumber\\
& &-a^2[rb'\mathfrak{L}^2(r\mathfrak{H}\mathfrak{H}'+\mathfrak{L}^2)-b(r^2\mathfrak{H}'^2-2\mathfrak{L}^2)]\Biggr]-2\mathfrak{L}^{3/2}[r^2ab(b+2)a''-r^2b(b+2)a'^2-raa'(r(b+4)b'+b^2+2b)\nonumber\\
& & -a^2(rb'(b+2)+2b^2+2b]\Biggr),\nonumber\\
& & S_{[2 0] 1} T^{,1}=\frac{\mathfrak{H}\sin\phi}{2\mathfrak{L}^{3/2}a b^2 r^3}\Biggl(b\Biggl[r^2ab \mathfrak{L}^4a''+r^2\mathfrak{H}a^2b \mathfrak{L}^4 \mathfrak{H}''-r^2b \mathfrak{L}^4a'^2-raa'\mathfrak{L}^2(r\mathfrak{L}^2b'-b [ r\mathfrak{H}\mathfrak{H}'-\mathfrak{L}^2] )\nonumber\\
& & -a^2[rb'\mathfrak{L}^2(r\mathfrak{H}\mathfrak{H}'+\mathfrak{L}^2)-b(r^2\mathfrak{H}'^2-2\mathfrak{L}^2)]\Biggr]-2\mathfrak{L}^{3/2}[r^2ab(b+2)a''-r^2b(b+2)a'^2-raa'(r(b+4)b'+b^2+2b)\nonumber\\
& & -a^2(rb'(b+2)+2b^2+2b]\Biggr).\end{eqnarray}
Solution of Eq. (44) has the form
\begin{equation}
\mathfrak{H}=\pm\frac{\sqrt{1-9r^4c_2{}^2\Lambda-27rc_2{}^2c_1}}{3c_2\sqrt{r(r^3\Lambda+3c_1)}}.\end{equation} Equation (45) is a solution of Eq. (44) and when we use it Eq. (41) we get  \begin{equation} E=M+\left(\frac{1}{r}\right).\end{equation} which is identical with Eq. (37).  This means that the solution of the skew-symmetric tensor removes the inertia from the true physics.
\newpage
\section{ Discussion and conclusion }

We have discussed the TEGR theory and its 6 extra degrees of freedom. For this purpose  we have studied three tetrad fields, with the flat horizon, reproduce  the same metric. The first tetrad field has two unknown functions and the field equations of TEGR give an analytic form of these functions. The conserved quantities of this tetrad are calculated and we have got a finite conserved quantity in the temporal coordinate  after using the Regularization using relocalization. We coined this tetrad as the physical tetrad. The reason for this name comes from the fact that we have defined a skew-symmetric tensor that is must vanish identically in the framework of TEGR. The first tetrad satisfied this property.

For the second tetrad which is obtained from the first one by multiplied it by a rotation matrix. We have calculated the scalar torsion of this tetrad and have shown that the scalar torsion is not invariant under local Lorentz transformation. We  have calculated the field equations and show that they are not different from the first tetrad and consequently the solution of the two unknown functions are the same as of the first tetrad. The discussion of the conserved quantities and the skew-symmetric tensors are the same of the first tetrad.

For the third tetrad, we have shown that it is related to the first tetrad through a local Lorentz transformation that contains an arbitrary function, $\mathfrak{H}$. We have calculated the scalar torsion of this tetrad and have shown that it depends on the arbitrary function. As usual, the field equations of TEGR can not fix any form of the arbitrary function. We have calculated the conserved quantities of this tetrad and have shown that the temporal component of the coordinate depends on the arbitrary function. This shows in  a clear way that the inertia contributes to the true physics. We have calculated the skew-symmetric tensor of this tetrad and have shown that some of its components are not vanishing. We have solved these non-vanishing components and have derived a form of the arbitrary function. When we have substituted this solution in the form of the conserved quantities we have shown that the energy will coincide with the value of the physical tetrad.
{\centerline{\bf Appendix A}}
{\centerline{\bf Notation}}
The indices ${\rm  i, j, \cdots }$ are the  (co)-frame components whilst  $\mu$, $\nu$,
$\cdots$ are the holonomic
spacetime coordinates. The  hats  $\hat{0}$,$\hat{1}$,  $\cdots$ c indicate special frame
components. The exterior product is denoted by $\wedge$.  The
interior  product of  $\xi$ and  $\Psi$ is described
by $\xi \rfloor \Psi$. The vector dual to the  1-forms
$\vartheta^{i}$ is labeled  by $e_i$ and their inner product satisfy
$e_i \rfloor {\rm \vartheta}^{j}={\delta_i}^j$. Employing
local coordinates $x^\mu$, we have ${\rm \vartheta}^{i}={e^i}_\mu
dx^\mu$ and $e_i={e_i}^\mu \partial_\mu$ where ${e^i}_\mu$ and
${e_i}^\mu $ are the components  covariant and contravariant of
the tetrad fields. The volume $\eta:=\vartheta^{\hat{0}}\wedge \vartheta^{\hat{1}}\wedge
\vartheta^{\hat{2}}\wedge\vartheta^{\hat{3}}$  defines  4-form.  Moreover,  we can altered
\[{\rm \eta}_i:=e_i \rfloor \eta = \ \frac{1}{3!} \
\epsilon_{i j k l} \ {\rm \vartheta}^j \wedge
{\rm \vartheta}^k \wedge {\rm \vartheta}^l,\]
where
$\epsilon_{ i j k l}$ is totally antisymmetric
with $\epsilon_{0123}=1$. \[\eta_{ i j}:=e_j \rfloor \eta_i =
\frac{1}{2!}\epsilon_{i j k l} \
{\rm \vartheta}^k \wedge {\rm \vartheta}^l,\qquad \qquad
\eta_{i j k}:=e_k
\rfloor \eta_{i j}= \frac{1}{1!} \epsilon_{i j k l} \ {\rm \vartheta}^l.\]  Finally,  we can define \[\eta_{i j k l}
:=e_l \rfloor \eta_{i j k}=
e_l \rfloor e_k \rfloor e_j \rfloor e_i \rfloor
\eta,\] which is the tensor density of Levi-Civita. The following useful identities\begin{eqnarray*} {\rm \vartheta}^i \wedge
\eta_j &:= & \delta^i_j
\eta, \qquad {\rm \vartheta}^i \wedge \eta_{j k}  := \delta^i_k \eta_j-\delta^i_j \eta_k,
\qquad  {\rm \vartheta}^i \wedge \eta_{j k l} := \delta^i_j \eta_{k l}+\delta^i_k
\eta_{l j}+\delta^i_l \eta_{ j k}, \nonumber\\
{\rm \vartheta}^i \wedge \eta_{j k l n}  & :=& \delta^i_n \eta_{j k l}-\delta^i_l \eta_{j k n
}+\delta^i_k \eta_{ j l n}-\delta^i_j
\eta_{k l n}\end{eqnarray*} are holds using the $\eta$-forms.

The line element $ds^2 :=g_{i j}
\vartheta^i \bigotimes \vartheta^j$ is defined by the
spacetime metric $g_{i j}$ .\vspace{0.3cm}\\
{\centerline{\bf Appendix B: Calculations of Weyl tensor and the object $W^{\mu \nu}$}}

The non-vanishing components of Weyl
tensor using solution of tetrad (10) have the form: \begin{eqnarray}& & C_{0101}=-C_{0110} =C_{1010}=-C_{1001}=\frac{-\sqrt{3}c_1c_2}{r^3},\nonumber\\
&&
C_{0202}=-C_{0220}=C_{0303}=-C_{0330}=C_{2020}=-C_{2002}=C_{3030}=-C_{3003}=\frac{-\sqrt{\Lambda r^3+3c_2}c_1c_2}{2r^{5/2}},\nonumber\\
& & C_{1212}=-C_{1221}=C_{1313}=-C_{1331}=-C_{2112}=C_{2121}=-C_{3113}=C_{3131}=\frac{-\sqrt{3}c_1}{2r^{3/2\sqrt{\Lambda r^3+3c_2}}},\nonumber\\
& & C_{2323}=-C_{2332}=-C_{3223}=C_{3232}=-\frac{c_1}{r}.\end{eqnarray}
The non-vanishing components of $W^{\mu \nu}$  are given by
\begin{eqnarray} & & W^{01}=-\frac{\sqrt{3}c_1c_2}{r^3}(dr \wedge dt)], \qquad W^{02}=-\frac{c_1c_2\sqrt{\Lambda r^3+3c_2}}{2r^{5/2}}(d\theta \wedge dt),\nonumber\\
& & W^{03}=-\frac{c_1c_2\sqrt{\Lambda r^3+3c_2}}{2r^{5/2}}(d\phi \wedge dt),\qquad W^{12}=-\frac{\sqrt{3}c_1}{2r^{3/2}\sqrt{\Lambda r^3+3c_2}}(dr \wedge d\theta),\nonumber\\
&  & W^{13}=-\frac{\sqrt{3}c_1}{2r^{3/2}\sqrt{\Lambda r^3+3c_2}}(dr \wedge d\phi), \qquad W^{23}=-\frac{c_1}{r}(d\theta \wedge d\phi).\nonumber\\
\end{eqnarray}
By the same method we can calculate the non-vanishing of Weyl tensor and of the object $W^{\mu \nu}$ of the second and third tetrad.

\end{document}